\documentclass[iop,floatfix]{emulateapj}
\pdfoutput=1

\RequirePackage{color}
\input{astro.sty}

\def\plotmode{col}

\def\plotext{pdf}
\newcommand\changed[1]{#1}

\def\picdir{.}

\shorttitle{The PS1 Photometric Reference Ladder}
\shortauthors{E.A. Magnier et al}
\begin{document}
\title{The Pan-STARRS 1 Photometric Reference Ladder, Release 12.01}

\def\IfA{1}
\def\CfA{2}
\def\MPIA{3}
\def\Princeton{4}

\author{
E.~A. Magnier,\altaffilmark{\IfA}
E. Schlafly,\altaffilmark{\CfA,\MPIA}
D. Finkbeiner,\altaffilmark{\CfA}
M. Juric,\altaffilmark{\CfA}
J.~L. Tonry, \altaffilmark{\IfA}
W.~S. Burgett,\altaffilmark{\IfA}
K.~C. Chambers,\altaffilmark{\IfA} 
H.~A. Flewelling,\altaffilmark{\IfA}
\changed{N. Kaiser,\altaffilmark{\IfA}}
R.-P. Kudritzki,\altaffilmark{\IfA}
J.~S. Morgan,\altaffilmark{\IfA}
P.~A. Price,\altaffilmark{\Princeton}
W.~E. Sweeney,\altaffilmark{\IfA}
C.~W. Stubbs,\altaffilmark{\CfA}
} 

\altaffiltext{\IfA}{Institute for Astronomy, University of Hawaii, 2680 Woodlawn Drive, Honolulu HI 96822}
\altaffiltext{\CfA}{Harvard-Smithsonian Center for Astrophysics, 60 Garden Street, Cambridge, MA 02138}
\altaffiltext{\MPIA}{Max Planck Institute for Astronomy, K\"onigstuhl 17, D-69117 Heidelberg, Germany} 
\altaffiltext{\Princeton}{Department of Astrophysical Sciences, Princeton University, Princeton, NJ 08544, USA}


\begin{abstract}

As of 2012 Jan 21, the Pan-STARRS\,1 $3\pi$ Survey has observed the
3/4 of the sky visible from Hawaii with a minimum of 2 and mean of 7.6
observations in 5 filters, \grizy.  Now at the end of the second
year of the mission, we are in a position to make an initial public
release of a portion of this unprecedented dataset.  
This article describes the PS1 Photometric Ladder, Release 12.01
This is the first of a series of data releases to be generated as the
survey coverage increases and the data analysis improves.  The
Photometric Ladder has rungs every hour in RA and at 4 intervals in
declination.  We will release updates with increased area coverage
(more rungs) from the latest dataset until the PS1 survey and the
final re-reduction are completed.  The currently released catalog
presents photometry of $\sim 1000$ objects per square degree in the
rungs of the ladder.  Saturation occurs at $\gps, \rps, \ips \sim 13.5$;
$\zps \sim 13.0$; and $\yps \sim 12.0$.  Photometry is provided for
stars down to $\gps, \rps, \ips \sim 19.1$ in the AB system.
\changed{This data release depends on the rigid `Ubercal' photometric
calibration using only the photometric nights, with systematic
uncertainties of (8.0, 7.0, 9.0, 10.7, 12.4) millimags in (\grizy).}
Areas covered only with lower quality nights are also included, and
have been tied to the Ubercal solution via relative photometry;
photometric accuracy of the non-photometric regions is lower and
should be used with caution.
\end{abstract}

\keywords{Surveys:\PSONE -- Techniques: photometric }

\vfil
\eject
\clearpage

\section{INTRODUCTION}
\label{sec:intro}

Accurate photometry is one of the key tools of astronomy, yet
ground-base astronomy continues to suffer the traditional challenge of
accurate photometric calibration.  Careful analysis of standard
ground-based CCD imaging data can yield relative photometry of objects
in a small field with accuracy of 1\% or better, and exotic techniques
can produce much higher accuracy in special cases
\cite[e.g.,][]{Tonry.2005,Johnson.2009}.  The remaining challenge for
the generic astronomer is that of calibration of the photometry of a
given field into one of the standard systems.

Part of the difficulty comes from the various effective bandpasses
due to the different telescopes and cameras used by observers.
Critical attention must be paid to the bandpass defined by the choice
of filter, detector, etc \citep[see, \eg][]{Bessel.1990}.  However,
for main sequence stars of spectral types earlier than $\sim$ K0, these bandpass
effects can be dealt with through measured color transformations.  The
larger challenge comes from the calibration of the images with respect
to a well-defined reference.

Calibration of images obtained from the ground is naturally difficult
due to the large area of the sky and the small sizes of detectors,
even modern large mosaic cameras.  Calibration of most individual
images still ultimately relies on observations of objects in reference
fields outside of the field of view of interest.  Since there is
currently no all-sky network of calibrated references with
sufficiently dense spacing, the calibration observations invariably do
not overlap the science fields.  Photometric calibration is thus
sensitive to the systematic impact of changes in the atmosphere or the
instrument between the observation of the science field and the
reference field.

Ground-based optical astronomy is on the verge of escaping this
traditional problem -- several projects are now (or will soon be) able
to define dense large-area networks of reference photometry with
accuracies at or below the 1\% level.  The Sloan Digital Sky Survey
\citep{SDSS.York}, for example, provides good photometry across the
$1.4\pi$ steradians covered by that survey (up through DR9), with
reported systematic errors of $\sim 10$ millimagnitudes for the
$g,r,i,z$ filters \citep{SDSS.Padmanabhan}.  LSST will eventually
provide the same or better for the $>2\pi$ steradians of the portion of the
sky covered by that survey.

The \PSONE\ project is surveying the $3\pi$ steradians north of $-30$
deg declination \cite{PS.MDRM}.  A major goal of this survey project
is the construction of a precision photometry reference catalog
covering the entire $3\pi$ region.  The design of the survey and the
careful attention to calibration concerns place it in an excellent
position to provide a photometric (and astrometric) reference for this
entire $3\pi$ region.  The \PSONE\ Survey is continuing, and final
calibration will not be complete for $\sim$ 6 months after the survey
completion, roughly mid 2014.  At this point in time, however,
interest in the community has grown strong in a preliminary data
release to demonstrate the progress at precision photometric
calibration, and to potentially provide reference data in the PS1
photometry system, and in regions not currently served by the SDSS
survey.

\section{Pan-STARRS1}
\label{sec.ps1}

\begin{table}
\caption{\PSONE\ $3\pi$ Survey Coverage Statistics\label{table:coverage}}
\tiny
\begin{center}
\begin{tabular}{l|rrr|l|rr}
\hline
\hline
& \multicolumn{3}{c|}{\bf Image coverage$^1$} & {\bf 50\%} & \multicolumn{2}{c}{\bf Exposure Coverage$^3$} \\
{\bf Filter} & {\bf 95\%} & {\bf 50\%} & {\bf 5\%} & {\bf Det$^2$} & {\bf 3\degree\ FOV} & {\bf 3.3\degree\ FOV} \\
\hline
\gps & 1.14 & 4.76 &  9.12 & 3.93 & 5.77 & 6.98 \\ 
\rps & 1.06 & 4.48 &  9.15 & 3.70 & 5.63 & 6.81 \\
\ips & 1.13 & 4.81 &  9.31 & 3.97 & 5.85 & 7.08 \\
\zps & 2.61 & 6.24 & 10.95 & 5.15 & 7.29 & 8.83 \\
\yps & 2.92 & 6.37 & 10.50 & 5.28 & 7.43 & 8.99 \\
\hline
\end{tabular}
\end{center}
$^1$(95,50,5)\% of the $3\pi$ Survey region is covered by the given number of images as of 2011 Jan 21. \\
$^2$Interpolated Median number of detections per filter for bright sources \\
$^3$Mean coverage expected for a filled circular focal plane with the given diameter \\
\end{table}

The 1.8m \PSONE\ telescope, located on the summit of Haleakala on the
island of Maui in the Hawaiian island chain, has been performing a set
of astronomical surveys since May 2010.  The long-term goals of the
Pan-STARRS project include the construction of a four telescope system
(Kaiser et al. 2002) on Mauna Kea; the \PSONE\ telescope, analysis,
and data publication system \cite{PS1.system} are prototypes of
the full hardware and software systems required for the 4 telescope
array.  \PSONE\ is currently being operated full-time in survey mode,
with several intertwined programs.  The surveys are designed around a
wide-ranging set of science drivers, addressing astronomical issues as
diverse as the contents of the inner solar system \changed{\citep[e.g.,
the discovery of the main belt comet Comet P/2006 VW139]{Hsieh-2012}} to galaxy clustering and cosmology using
Type Ia supernovae \citep[e.g.,][]{2011ApJ...731L..11N}.

The wide-field optical design of the \PSONE\ telescope
\citep{PS1.optics}, based on a 1.8~meter diameter $f$/4.4 primary
mirror and an 0.9~m secondary, produces a 3.3 degree field of view
with low distortion and minimal vignetting even at the edges of the
illuminated region.  The optics, in combination with the natural
seeing, result in generally good image quality: 75\% of the images
have full-width half-max values less than (1.51, 1.39, 1.34, 1.27,
1.21) arcseconds for (\grizy), with a floor of $\sim 0.7$ arcseconds.
The \PSONE\ camera \citep{PS1.GPCA} consists of a mosaic of 60
edge-abutted $4800\times4800$ pixel detectors, with 10~$\mu$m pixels
subtending 0.258~arcsec.  The detectors are back-illuminated CCDs
manufactured by Lincoln Laboratory and are read out using a StarGrasp
CCD controller, with a readout time of 7 seconds for a full unbinned
image. Initial performance assessments are presented in
\cite{PS1.GPCB}.  The active, usable pixels cover $\sim 80$\% of the
FOV.  Routine observations are conducted remotely from the Advanced
Technology Research Center in Kula, the main facility of the
University of Hawaii's Institute for Astronomy operations on Maui.

\begin{figure*}[htbp]
\begin{center}
\includegraphics[type=\plotext,ext=.\plotext,read=.\plotext,width=6.2in]{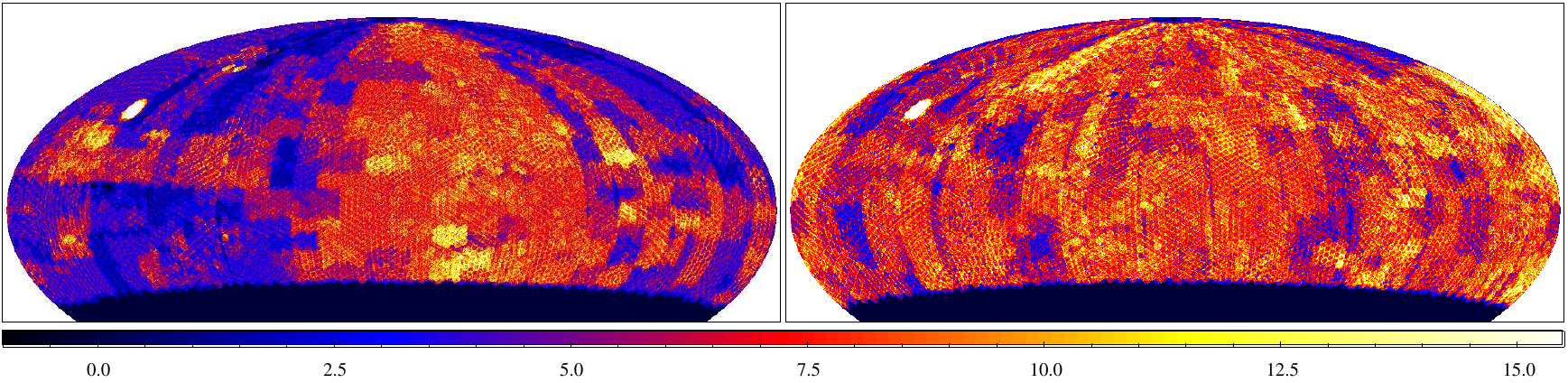}
\caption{Plots of the sky coverage of the \TPS\ through 21 Jan 2012.
  The color (greyscale in print version) shows the number of separate
  exposures overlapping the given location, as indicated by the scale.
  The left panel shows the distribution for \gps\ while the right
  panel shows the distribution for \yps.  The distributions for
  \rps\ \& \ips\ are similar to \gps, while that of \zps\ mimics \yps.
  RA = 0.0 is at the center of the plots and increases to the left.
  The patch at $09^h 15^m, +30\degree 45\arcmin$ was observed with the
  full set of $3\pi$ exposures in 2010 Feb as a demonstration
  dataset.} \label{fig:coverage}
\end{center}
\end{figure*}

Images obtained by the \PSONE\ system are saved and processed on a
dedicated data analysis cluster located at the Maui High Performance
Computer Center in Kihei, Maui.  Observations are automatically
processed in real time by the \PSONE\ Image Processing Pipeline
\citep[IPP,][]{PS1.IPP}, with the ultimate goals being the 1) characterization
of astronomical objects in the individual images; 2) construction of
stacks of multiple images of the same areas of the sky to improve the
sensitivity and to fill in gaps (along with the characterization of
the objects in those stacks); 3) construction of difference images
between individual images or stacks and reference images from another
epoch for the purpose of detecting variable and moving objects.  This
article relies only on data from the analysis of the individual
exposures, not on any of the stacked or difference images.

In more detail, individual images are detrended: non-linearity and
bias corrections are applied, a dark current model is subtracted and
flat-field corrections are applied.  The \yps-band images are also
corrected for fringing: a master fringe pattern is scaled to match the
observed fringing and subtracted.  Mask and variance image arrays are
generated with the \changed{detrend analysis} and carried forward at each stage
of the IPP processing.  Source detection and photometry are performed
for each chip independently.  Astrometric and photometric calibrations
are performed for all chips together in a single exposure.

For the first 20 months, lacking all-sky photometry in the
\PS1\ band-passes, a provisional photometric calibration was performed
on the nightly exposures using a synthetic photometric catalog based
on the combined fluxes in Tycho, USNO-B, and 2MASS.  In practice,
since Tycho only contributes very bright stars, and the large
photometric errors for USNO stars mean they carry little weight, these
calibrations are effectively tied to 2MASS as if all stars were on the
main sequence color locus.  The resulting photometric calibrations are
observed to be no better than $\sim 5$\%.

Global re-calibration analysis of the photometric \PSONE\ data, as
discussed in more detail below, has been used to generate a PS1-based
reference catalog which is now used for nightly science photometric
calibration \citep{PS1.ubercal}.  This reference catalog also uses the
internally improved relative astrometric calibration for improved
\changed{accuracy of the astrometry}.

\subsection{The \PSONE\  \TPS}

The PS1 telescope is operated by the PS1 Science Consortium to perform
a set of inter-twined surveys with a range of spatial and temporal
coverage.  Several narrow-field ``Medium Deep'' surveys are
complemented by the wide-area \TPS, the latter allocated 56\% of the
available observing time.  The \TPS\ aims to observe the portion of
the sky North of $-30$ deg declination, with a total of 20 exposures
per year in all filters for each field center.  The \TPS\ observations
are performed with extensive dithering \changed{between different exposures} so
that the overlaps can be used to tie down the photometric and
astrometric system.

The \TPS\ observations are performed using the five main filters
(\grizy).  Details of the passband shapes are provided by
\cite{JTphoto}.  Provisional response functions (including 1.2
airmasses of atmosphere) are available at the project's web
site\footnote[1]{http://svn.pan-starrs.ifa.hawaii.edu/trac/ipp/wiki/PS1\_Photometric\_System}.
The full survey strategy is described in \cite{PS.MDRM}; we summarize
the salient details below.


\subsubsection{Observing Strategy}

The \TPS\ observations are performed on a complex schedule in order to
balance the needs of the different survey science projects.  The goal is
to obtain a total of 12 observations in each filter for any spot in
the observable region by the end of the survey mission.  This goal
actually refers to a theoretical survey with a perfectly filled focal
plane with no gaps or overlaps between neighboring observations.  In
practice, gaps in the focal plane (between cells, between chips, and
from masked pixels) result in a fill-factor of $\sim 80$\% for a
single exposure.  Neighboring exposures have both overlapping areas as
well as gaps between the exposure due to the layout of the chips in
the camera.  The net effect of the overlaps and incomplete fill factor
is illustrated in Table~\ref{table:coverage}, discussed in more detail
below.  While the goal is to achieve the full $3\pi$ coverage in 3
years of observations, it is expected that an additional 6 - 12 months
of cleanup observations will be needed to fill in holes due to weather
and other down-time, and to improve the homogeneity of the full
survey.

The temporal distribution of the 12 observations per filter is
somewhat complicated.  The following guidelines are used by the
observing system, though some observations will inevitably fail to
meet these goals.  First, any specific field is always observed 2
times in a single night in a single filter, nominally within 20 - 30
minutes.  This so-called ``Transient Time Interval'' allow for the
discovery of moving objects (asteroids and NEOs); as part of the
nightly processing these ``TTI-pairs'' are mutually subtracted and
objects detected in the difference image are reported to the Moving
Object Pipeline Software (MOPS).  Second, the blue bands (\gps, \rps, \ips)
are observed close to opposition to enable asteroid discovery.  These
observations normally occur within $\sim 1.5$ months of opposition for
any given field.  Thus any given field should be observed a total of
12 times in these 3 filters within a 2-3 month window each year.  For
the reddest 2 bands (\zps\ \& \yps), the observations are scheduled as
far from opposition as feasible in order to enhance the parallax
factors and allow for discovery of faint, low-mass objects in the
solar neighborhood.  This constraint results in 2 observations in each
of \zps\ \& \yps\ occurring roughly 4-6 months before and 4-6 months
after opposition for any given field. 

The pointing of individual observations is designed to trade off
between maximal overlaps and optimized image differencing.  The TTI
pair images are, as much as possible, obtained in the same pointing,
both location of the boresight and rotation on the sky, in order to
minimize the loss of area in the difference image from mis-matched
gaps.  Subsequent sets of TTI pairs are obtained at offset positions
to tile over the boundaries between neighboring exposures.  A set of
telescope pointings are defined to carefully cover the entire $3\pi$
region with a judicious balance between the total number of exposures
and the total area missed due to gaps between the exposures.  One set
of these telescope pointings is called a ``$3\pi$ tessellation''.  The
base tessellation is used for the first TTI pair of observations.  
Subsequent observations follow a $3\pi$ tessellation offset from the base
by Euler angle rotations which shift most of the boresights by $\sim
50\%$ of the field diameter.  Note that no single pair of
tessellations can have all field centers offset by a fixed amount;
for any pair of tessellations, some subset of boresight positions will
have only small offsets.  By choosing different poles for these
rotations, all fields can have most observations with substantial
offsets.  

The end result of the scheduling and dithering strategy is that the
full $3\pi$ region is covered in a wide range of time periods in each
filter and has a large range of spatial overlaps.  These overlaps
provide a tight mesh for any global photometric and astrometric
solution, while the temporal coverage increases the chances that all
areas receive photometric observations, or are at least not too
distant from photometric data.

\subsubsection{Survey Coverage to Date}

\changed{The scheduling constraints described above, when coupled with vagaries
of weather, the need to avoid the moon, and the other (non-3$\pi$)
survey observations make the actual scheduling of the observations
quite challenging.  The overall observing strategy on a nightly basis
is to start and end the nights with \zps\ or \yps\ observations and to
schedule the \gps, \rps\ and \ips\ observations for the hours near
midnight.  The 3$\pi$ observation blocks are interspersed with
observations for the other survey components (e.g., Medium Deep
fields).  In addition, as the lunation progresses, the amount of time
used for the red or blue bands is shifted as the sky brightness
increases.  In general, observations are obtained fairly close to the
Meridian: for observations away from the zenith keyhole, the
hour-angle distribution is Gaussian with $\sigma \sim 1$ hour.  }

Figure~\ref{fig:coverage} illustrates the spatial coverage of the
\TPS\ to date (2012/01/21).  The left panel shows the density of
coverage for \gps, while the right panel shows the density of coverage
for \yps.  The blue bands (\gps, \rps, \ips) have generally similar
coverage, as do the two red bands.  To date, the blue bands have been
somewhat more affected by weather.  The median coverage for any point
in the $3\pi$ region is between 4.5 and 6.3 exposures, depending on
the band.  \changed{With $\sim 5$ exposures on average in $\sim 1.5$ years of
survey operations, we are generally on track to achieve 12 visits by
the end of the 3.5 year mission.  However, while some effort is being
made in the second half of the survey to fill in gaps due to the
weather, the final coverage will likely not be extremely uniform.  The
top-level goal is to obtain photometric observations for all areas in
all filters, though this will be challenging.}

These coverage numbers are defined based on the overlap of
chip boundaries for those chips actually loaded into the photometry
database.  The median coverage in the $3\pi$ region is defined as the
number of chips covering 50\% of patches in the full $3\pi$ region.
Table~\ref{table:coverage} lists the median coverage for each filter
as well as the 5\% and 95\% coverage values.  Since there are dead
spaces in chips (gaps between cells and masked pixels), the number of
detections for a well-measured star will be somewhat less than the
number of overlapping images.  This effect is reflected in the 5th
column of Table~\ref{table:coverage}, which gives the median number of
detections for moderately bright (S/N $\sim$ 50) objects.  The last
two columns of the table give the expected coverage of the full $3\pi$
region (down to Dec of -32\degree, where we have partial coverage), if
PS1 had a completely filled circular focal plane with field of view
3.0 and 3.3\degree.


The database includes measurements with all quality levels.  The data
processing version used for this analysis has suffered from a
relatively high rate of artifacts (false positive detections), either
due to instrumental structures, optical features such as ghosts and
diffraction spikes, and background subtraction failures.  Since
January 2012, the image processing team have made substantial
improvements in reducing these sources of false positives, and in
flagging the artifacts which remain.  However, the data used in this
work have not yet had the benefit of these improvements.  For the
purposes of this article, we can robustly filter out these false
detections by requiring multiple observations of the same object and
by restricting our attention to brighter objects.  The database used
for this analysis contains $3.03 \times 10^{10}$ measurements of $2.2
\times 10^9$ objects at all magnitudes; including only those objects
with 3 or more detections, the dataset includes $2.66 \times 10^{10}$
detections of $1.49 \times 10^9$ objects.

\subsubsection{Overall System Zero Points}

\begin{figure*}[htbp]
\begin{center}
\includegraphics[type=\plotext,ext=.\plotext,read=.\plotext,width=6.5in]{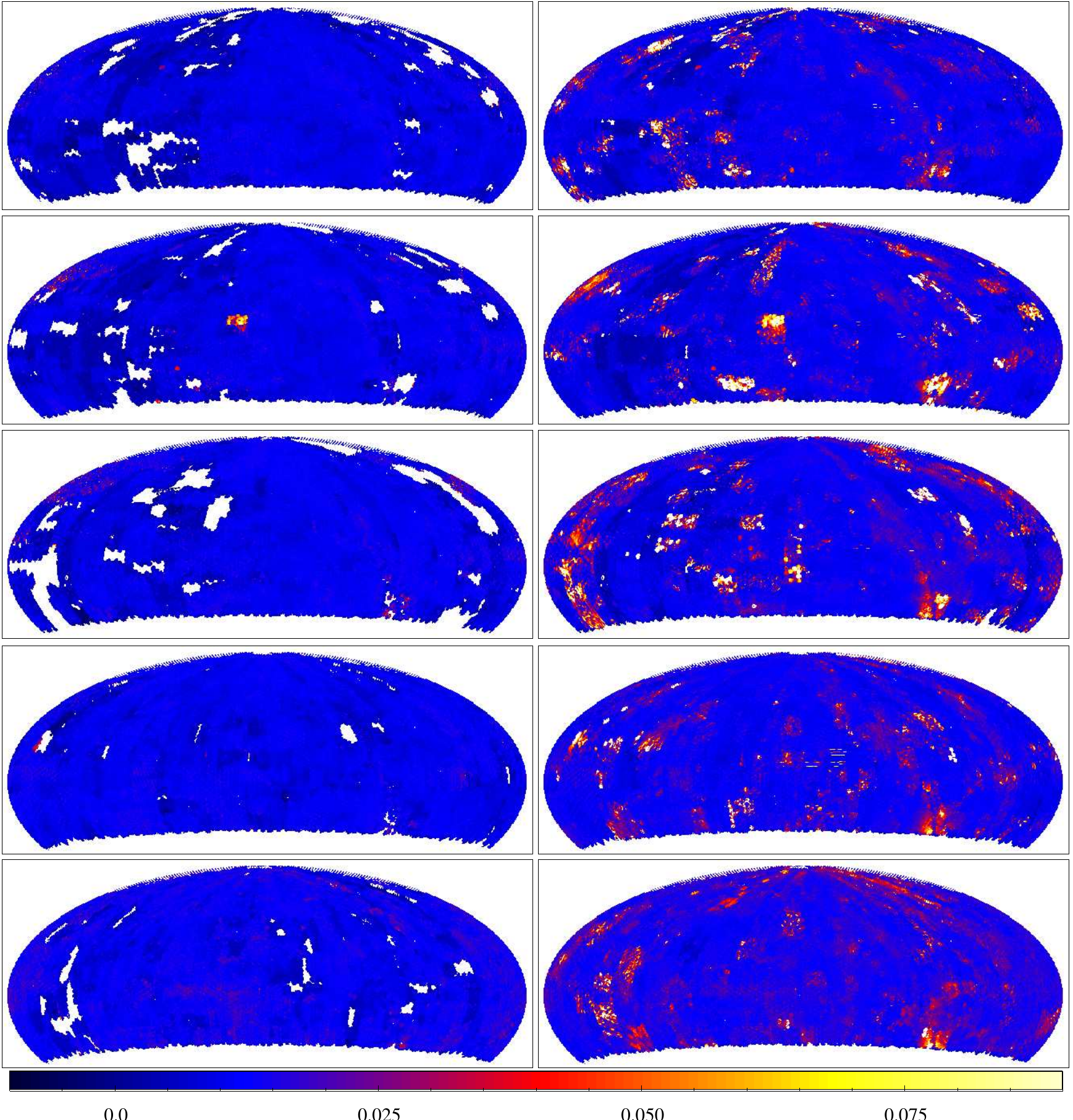}
\caption{Plots of the scatter for bright stars using just Ubercal data
  (left) or all PS1 measurements (right).  The color scale shows the
  amplitude of the scatter in magnitudes.  RA = 0.0 is at the center
  of the plots and increases to the left.} \label{fig:sigmas}
\end{center}
\end{figure*}

Traditional astronomical magnitudes, such as the Johnson-Kron-Cousins
$UBVRI$ system, have been defined as 2.5 times the logarithm of the
ratio of fluxes between the object of interest as observed with the
given telescope to that of the star Vega observed with the same
instrumentation.  Some of the drawbacks of this system include (1) a
strong dependence on the quality of our knowledge of the magnitudes of
Vega in any bandpass of interest and (2) the difficulty of observing
one of the brightest stars in the sky with the same instrument used to
observe some of the faintest objects known.

\PSONE\ uses the alternative ``AB magnitude system'', introduced by
Oke \& Gunn 1983, in which the magnitude of an object is defined by
the integral of the flux density spectrum multiplied by the overall
system throughput as a function of wavelength for the telescope of
interest.  Symbolically, for a telescope with a system response of
$A(\nu)$ and an object with a flux density spectrum of $f_\nu$
(erg/sec/cm$^2$/Hz), the AB magnitude for a band-pass is defined to
be:
\begin{equation} \label{eq.mab}
m_{AB} = -2.5 \log \frac{\int f_\nu (h\nu)^{-1} A(\nu) d\nu}{\int 3631 \mbox{Jy} (h\nu)^{-1} A(\nu) d\nu}
\end{equation}
While this system is not subject to the drawbacks suffered by the Vega
system, it has difficulties of its own.  In particular, the accuracy
of the calibration for any given telescope is limited by (1) our
knowledge of the system response (including the atmosphere!) and (2)
our knowledge of the spectral energy distribution of a specific star
of interest.  In practice, like the Vega system, broad-band magnitudes
are defined by comparison with the observed magnitudes of well-known
stars.  Since stars with spectral types earlier than $\sim$ early K
are predominantly a continuum source with minor absorption lines, the
broad-band colors of stars vary smoothly.  Unlike traditional
Vega-based photometry, however, in the AB system stars of colors
different from the spectrophotometric reference have well-defined
magnitudes as long as the bandpass is well measured.

\cite{JTphoto} have determined the overall system zero points
necessary to place the \PSONE\ magnitudes onto the AB system.  We
summarize the details of that analysis here.

First, they have
determined the relative spectral response of the \PSONE\ system from
the top of the telescope to the detector, in the absence of filters,
using a tunable laser and a NIST-calibrated photodiode.  To do this,
the light from the tunable laser was used to illuminate both the pupil
of the telescope and the photodiode.  The GPC1 camera was used to
record the flux from the laser after it had passed though the full
optical system, with the filter holder in the `open' slot.  A sequence
of measurements was made for 2nm steps from 400 nm to 1100nm, and the
flux observed by the calibrated photodiode compared to the flux
observed by GPC1.  The resulting scans defined the relative response
of the system as a function of wavelength without the filters.

The filter transmission curves were measured by the manufacturer, Barr
Precision Optics, at a number of positions and angles of incidence.
\cite{JTphoto} also measured the filter curves including the filter in
the beam and running the tunable laser + diode system described above.
The two sets of data agree well; in their analysis \cite{JTphoto}
adopt the Barr curves as the reference set. 


The third optical element in the PS1 system is the atmosphere above
the telescope.  \cite{JTphoto} calculate the transmission of the
atmosphere as a function of wavelength with the model atmosphere
program MODTRAN \citep{MODTRAN}.  These data are used to calculate the
atmospheric extinction as a function of airmass, precipitable water
vapor (PWV), and the power-law slope of a given source SED.  The PS1
system has been monitoring the PWV content of the atmosphere since
June 2011.  

Putting together the above components, \cite{JTphoto} tie down the
overall system zero points by observing a selection of
spectrophotometric standards with PS1 in a photometric night.  The
above components of the spectral response of the system were combined
and applied to the reported SEDs of the 7 standards to predict
observed PS1 fluxes.  In addition, a large number of stars with
measured spectra were used to construct PS1 stellar locus diagrams to
provide an additional set of constraints.  Comparison of the predicted
and observed magnitudes in each of the 7 available PS1 filters
(\grizy, \wps, and open) led \cite{JTphoto} to introduce tweaks to 12
system parameters to obtain the best match.  These tweaks are all at
the $\sim$ 1\% level, and are the result of the accuracy achieved in
the spectral response measurement.

\subsubsection{The Ubercal Analysis}


\begin{table*}
\caption{Bit-flags used to exclude bad or low-quality detections} 
\begin{center} \tiny
\begin{tabular}{l|l|l|l}
\hline
\hline
{\bf FLAG NAME} & {\bf Hex Value} & {\bf Bad / Poor} & {\bf Description} \\
PM\_SOURCE\_MODE\_FAIL             	  & 0x00000008  &  Bad  & Fit (non-linear) failed (non-converge, off-edge, run to zero) \\
PM\_SOURCE\_MODE\_POOR             	  & 0x00000010  &  Poor & Fit succeeds, but low-SN or high-Chisq \\
PM\_SOURCE\_MODE\_PAIR             	  & 0x00000020  &  Poor & Source fitted with a double psf \\
PM\_SOURCE\_MODE\_SATSTAR          	  & 0x00000080  &  Bad  & Source model peak is above saturation \\
PM\_SOURCE\_MODE\_BLEND            	  & 0x00000100  &  Poor & Source is a blend with other sources \\
PM\_SOURCE\_MODE\_BADPSF           	  & 0x00000400  &  Bad  & Failed to get good estimate of object's PSF \\
PM\_SOURCE\_MODE\_DEFECT           	  & 0x00000800  &  Bad  & Source is thought to be a defect \\
PM\_SOURCE\_MODE\_SATURATED        	  & 0x00001000  &  Bad  & Source is thought to be saturated pixels (bleed trail) \\
PM\_SOURCE\_MODE\_CR\_LIMIT         	  & 0x00002000  &  Bad  & Source has crNsigma above limit \\
PM\_SOURCE\_MODE\_MOMENTS\_FAILURE  	  & 0x00008000  &  Bad  & could not measure the moments \\
PM\_SOURCE\_MODE\_SKY\_FAILURE      	  & 0x00010000  &  Bad  & could not measure the local sky \\
PM\_SOURCE\_MODE\_SKYVAR\_FAILURE   	  & 0x00020000  &  Bad  & could not measure the local sky variance \\
PM\_SOURCE\_MODE\_BELOW\_MOMENTS\_SN 	  & 0x00040000  &  Poor & moments not measured due to low S/N \\
PM\_SOURCE\_MODE\_BLEND\_FIT        	  & 0x00400000  &  Poor & source was fitted as a blend \\
PM\_SOURCE\_MODE\_SIZE\_SKIPPED     	  & 0x10000000  &  Bad  & size could not be determined \\
PM\_SOURCE\_MODE\_ON\_SPIKE         	  & 0x20000000  &  Poor & peak lands on diffraction spike \\
PM\_SOURCE\_MODE\_ON\_GHOST         	  & 0x40000000  &  Poor & peak lands on ghost or glint \\
PM\_SOURCE\_MODE\_OFF\_CHIP         	  & 0x80000000  &  Poor & peak lands off edge of chip \\
\hline
\end{tabular}
\end{center}
\label{table:photflags}
\end{table*}

\cite{PS1.ubercal} have reported on the photometric calibration of the
first 1.5 years of the PS1 survey.  In that analysis (``Ubercal''),
the photometric nights are selected and all other data are ignored.
Each night is allowed to have a single fitted zero point and a single
fitted value for the airmass extinction coefficient per filter (see
Eq.~\ref{eq.mcal} below).  Zero points of each night are determined by
minimizing the dispersion of the measurements of the same stars from
multiple nights.  \cite{PS1.ubercal} also determine flat-field
corrections as part of the minimization process.  The flat-field
corrections are measured for $2\times 2$ sub-regions of each chip in
the camera, and are determined by choosing zero point offsets for
these patches to minimize the scatter per star.  Four distinct time
periods (``seasons'') were identified in which these flat-field
corrections were quite consistent, but noticeably different from the
other seasons.  The transitions between the seasons have been
identified with specific changes to the optical system: modification
of the baffling structures and changes to the collimation and
alignment coefficients.  The underlying cause of the different
flat-fields is believed to be due to small scale changes in the
vignetting and PSF structure.

By excluding non-photometric data (both manually up front and
iteratively in the analysis) and only fitting 2 additional parameters
for each night, the Ubercal solution is both robust and extremely
rigid.  It is not subject to unexpected drift or sensitivity of the
solution to the vagaries of the data set.  The Ubercal analysis is
also especially aided by the inclusion of multiple Medium Deep field
observations every night, helping to tie down overall variations of
the system throughput and acting as internal standard star fields.
The resulting photometric system is shown by \cite{PS1.ubercal} to
have reliability across the survey region at the level of (8.0, 7.0,
9.0, 10.7, 12.4) millimags in (\grizy).  In addition, the consistency
of the measured zero points (scatter of $\sim 4$ millimag) hints at
the possibility of even better overall photometry as more information
is used to determine the flat-field variations.

The Ubercal analysis, performing a highly-constrained relative
photometry calculation, requires the external definition of the zero
point for each filter.  \cite{PS1.ubercal} set the zero points of their
images to match the values resulting from the analysis of
\cite{JTphoto}.  This is done by matching the photometry of the MD09
Medium Deep field to that measured by \cite{JTphoto} on the reference
photometric night of MJD 55744 (UT 02 July 2011).

\subsubsection{Relative Photometry}
\label{sec.relphot}

In this work we have used the Ubercal solution as a starting point,
and have used relative photometry between individual exposures to
determine the zero points for the data not calibrated by the Ubercal
analysis.  The combination of overlaps and the rigid base of the
Ubercal analysis allows us to determine reliable photometry for areas
which were excluded by \cite{PS1.ubercal}.  

The basic analysis is similar in many respects to the Ubercal
approach.  We start with a database of all observations, using the
Pan-STARRS Desktop Virtual Observatory software \citep[DVO,][]{PS1.IPP}.
The database links the table of individual `detections' to the images
from which they came, and groups the detections into unique
astronomical `objects'.  

Individual detections are characterized by a total number of counts
observed for the source, and converted to an instrumental magnitude:
$m^{inst} = -2.5\log_{10} (\mbox{counts}) + 2.5\log(\mbox{exptime})$.  The detected
objects (at least those which are not detectably variable) have an
intrinsic mean magnitude in the AB photometric system,
$m^{AB}$, of which each detection is a realization.  The instrumental
magnitude and the mean magnitude are related by an arithmetic offset
which accounts for various effects (zero point, instrumental
variations, atmospheric attenuation):
\begin{equation} \label{eq.msys} m^{AB} = m^{inst} + Z \end{equation} 
We can decompose the zero point $Z$ into the primary contributors as
follows:
\begin{equation} \label{eq.mcal} Z_{i,n} = a_n - k_n x_i + C_n * \mbox{color} + T_i \end{equation},
where $a_n$ and $k_n$ are the system zero point and trend with airmass
for a given period of time, while $x_i$ is the airmass ($\sec z$, for
zenith angle $z$) of a given image $i$.  We also allow for a trend
that depends on the $\mbox{color}$ of a star and an additional term
measuring any additional reduction in the transparency, $T_i$ for any
given image (e.g., clouds or haze).

In our analysis, we are taking the system zero point and the mean
airmass extinction coefficients provided by the Ubercal analysis, and
solving for a single additional offset for each exposure not already
tied down by the Ubercal analysis.  For this analysis, we neglect the
color difference of the different chips, and thus set the value of
$C_n$ to 0.0.  Note that we only use a single mean airmass extinction
term for all exposures -- the difference between the mean and the
specific value for a given night is taken up as an additional element
of the atmospheric attenuation.



We minimize the following global $\chi^2$ equation by finding the best
mean magnitudes for all objects and the best cloud offset for each
exposure:
\begin{equation} \label{eq.chisq} \chi^2 = \sum_{i,j} (m^{inst}_{i,j} + a_n - k_n x_i + T_i - m^{sys}_j) w_{i,j} / \sum_{i,j} w_{i,j} \end{equation}
where $i$ is the index for each image and $j$ is the index for each
star.  We set the weighting values $w_{i,j}$ to the inverse variance
of the individual measurements.  If everything were fitted at once and
allowed to float, this system of equations would have $N_{images} +
N_{stars} \sim 2 \times 10^5 + 2 \times 10^9$ unknowns.  We solve the
system of equations by iteration, solving first for the best set of
mean magnitudes in the assumption of zero clouds, then solving for the
clouds implied by the differences from these mean magnitudes.  Even
with 1-2 magnitudes of extinction, the offsets converge to the
milli-magnitude level within 8 iterations.  


After a series of 8 initial iterations, we perform outlier rejections.
For each star, the inner 50\% of measurements are used to define a
measurement of the standard deviation which is robust against
significant outliers.  Using this measurement of the standard
deviation, any measurements more than 5$\sigma$ deviant from the
median are excluded, and the mean \& standard deviation (weighted by
the inverse error) are recalculated.  The resulting values are used to
exclude detections which are more than 3$\sigma$ deviant from the
mean.  These deviant measurements are then flagged and excluded from
the rest of the analysis. 


Suspicious images and suspicious stars are also rejected.  For the
stars, we reject objects with $\chi^2_\nu$ values more than 20.0, or more
than 2$\times$ the median $\chi^2_\nu$ value, whichever of these cuts is
larger.  We also reject stars with scatter (standard deviation of the
measurements used for the mean) greater than 0.005 mags or 2$\times$
the median scatter, whichever is greater.  Similarly for images, we
reject those with more than 2 magnitudes of extinction or with scatter
greater than 0.075 mags or 2$\times$ the median scatter, whichever is
greater.  If the star \& detection rejection steps have the effect of
eliminating too many measurements for a given image, we exclude that
entire image.  The total number of valid measurements for an image
must be $> 10$ and the fraction of valid measurements to the total
number of measurements considered per image must be $> 5\%$.

These cuts are somewhat conservative to limit us to only
good measurements.  These images and stars are excluded when solving for
the system of zero points and mean magnitudes.  These cuts are updated
several times as the iterations proceed.  After the iterations have
completed, the poor-quality images are then calibrated based on their
overlaps with other images, and mean magnitudes for all stars are
calculated.

In this analysis, we use the Ubercal measurements as a rigid base by
setting the weight of Ubercal detections to 10x their default
(inverse-variance) weight.  The calculation of the formal error on the
mean magnitudes propagates this additional weight, so that the errors
on the Ubercal observations dominates where they are present:
\begin{equation} \mu = \frac{\sum m_i w_i \sigma_i^{-2}}{\sum w_i \sigma_i^{-2}} \end{equation}
\begin{equation} \sigma_\mu = \frac{\sum w_i^2 \sigma_i^{-2}}{(\sum w_i \sigma_i^{-2})^2} \end{equation}
where $w_i$ = 1.0 for non-Ubercal measurements and 10.0 for Ubercal measurements.

In practice, we further restrict the data used in the analysis.  We
use only the brighter objects, limiting the density to a maximum of
2500 or 3000 objects per square degree (lower in areas where we have
more observations).  When limiting the density, we prefer objects
which are brighter (but not saturated), and those with the most
measurements (to ensure better coverage over the available images).
We also exclude in advance those measurements for which the
photometric analysis flagged the result as suspicious (see Table~\ref{table:photflags}).  The latter
includes detections which are excessively affected by masks ({\tt
  PSF\_QF} $<$ 0.85), which land too close to other bright objects,
which are suspiciously close to diffraction spikes or ghost images, or
which are too close to the detector edges.

We perform the relative photometry analysis on a large area of the
sky, but not the entire database, in a given pass.  The driver here is
to avoid over-filling the memory of our analysis machine (48GB).
Another practical consideration is the data I/O needed for a given
analysis -- processing more, smaller areas costs more time for I/O
operations.  We have found that we achieve a good balance by splitting
the sky into 15 regions: 12 RA bands from -45\degree\ to +60\degree,
plus the polar region, with the 2 RA bands covering the Galactic
Center split in two at Dec = 5\degree.  An exposure must be completely
contained within an analysis region to have its zero point (cloud
offset) determined; images overlapping the edges of the analysis
region contribute measurements, but their offsets are held fixed for
that analysis region.  The analysis regions are defined to have 5
degree overlapping boundaries to ensure that all exposures have the
chance to be fitted.

\begin{figure*}[htbp]
\begin{center}
\parbox{6.5in}{
 \includegraphics[type=\plotext,ext=.\plotext,read=.\plotext,width=6.5in]{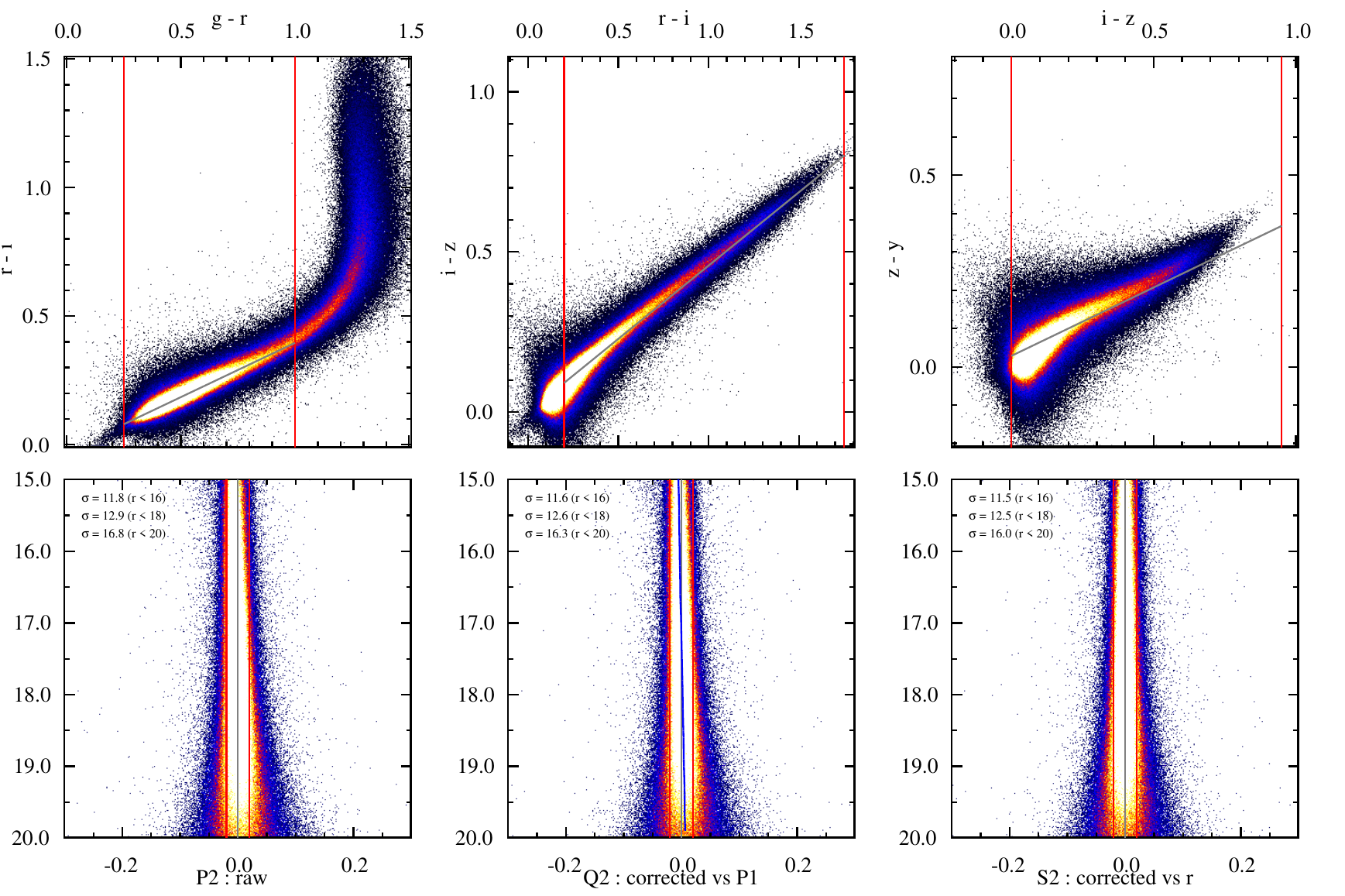}
}
\caption{Demonstration of the stellar locus fitting.  Top Row: the 3
  2-color planes used for this analysis.  The vertical (red) lines
  mark the range of colors used to define a stellar locus segment for
  each 2-color plane.  The grey line show the initial linear fit to
  this portion of the stellar locus.  Bottom Left: \rps\ vs the
  principal color P2 for the $\gps - \rps, \rps - \ips$ 2-color plane.
   Bottom Middle: \rps\ vs the Q2, the principal color after
   correction for variation of the stellar locus position as a
   function of the P1 dimension.  Bottom Right: \rps\ vs the S2, the principal color after
   correction for variation of the stellar locus position as a
   function of \rps.  The inset numbers show the measured scatter for
   stars with three faint limits.  Note the consistent improvement of
   the scatter for brighter objects.} \label{fig:stlocus.demo}
\end{center}
\end{figure*}

Once the image zero points \& cloud extinctions have been calculated,
these values are then applied to the individual measurements and final
mean magnitudes are calculated for all objects in the database.  Since
we are using the reference catalog within the Pan-STARRS project as
our internal calibration system, we would like to determine the best
set of magnitudes for all objects, regardless of the quality of the
data available for each object.  Even if we only have poor
measurements or no measurements of a given star, we would like to
populate the catalog with our best guess given the information
available.  To this end, we attempt to determine the mean magnitudes
in a series of passes.  In the first pass, we have a very conservative
selection of data to be used for the analysis.  In successive passes,
we relax our criteria for those objects which had no valid data in the
earlier passes.  The 5 different levels of data acceptance are:
\begin{itemize}
\item pass 0: only `good' measurements as reported by the photometry
  analysis; reject outliers determined by the relative photometry
  analysis
\item pass 1: accept measurements deemed `poor' by the photometry
  analysis (`suspect', but not `bad' masks; poor fit but not failed
  fit; etc.)
\item pass 2: accept the relative photometry outliers
\item pass 3: accept measurements deemed `bad' by the photometry
  analysis (eg, saturated stars, partially masked detections)
\item pass 4: accept 2MASS/Tycho/USNO-based synthetic photometry (see
  Section~\ref{sec.ps1} above).
\end{itemize}
For any object in the database, the mean magnitudes in the 5 filters
are independently tested for these different levels.  Which data
quality was used can be determined by examining the flags for that
filter.

A final note regarding the use of synthetic photometry: we have found
that very bright stars in the PS1 data can be split into multiple
measurements by the psphot analysis.  In these cases, the synthetic
photometry can be more reliable than the PS1 magnitudes.  We currently
accept the synthetic magnitudes if the implied \TPS\ magnitude would
be severely saturated (instrumental magnitudes $< -15$).

\subsubsection{Data Quality Checks}

\begin{figure*}[htbp]
\begin{center}
\includegraphics[type=\plotext,ext=.\plotext,read=.\plotext,width=6.0in]{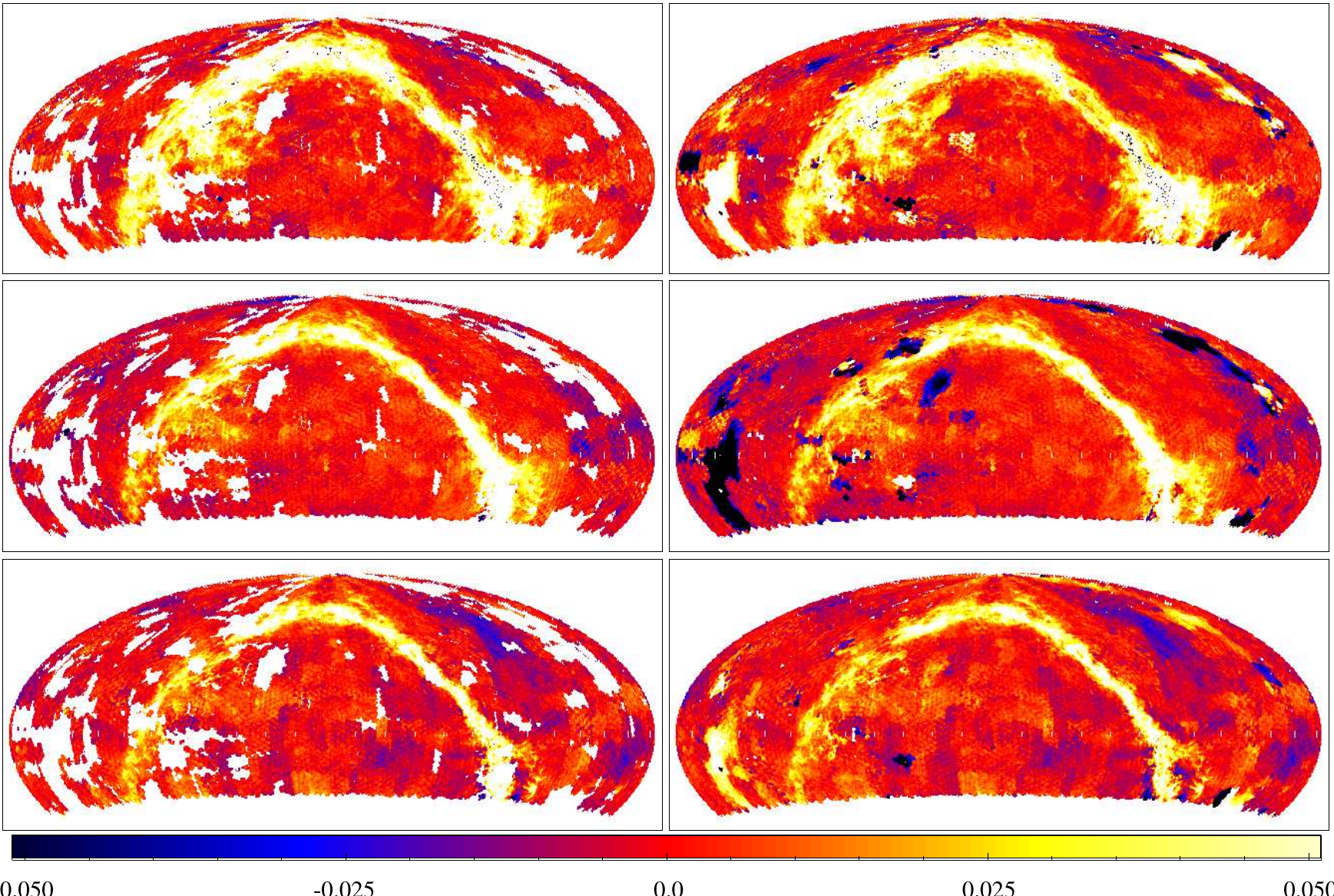}
\caption{Plots of the mean offset of the stellar locus for the 3
  2-color planes (top-to-bottom: $\gps - \rps, \rps - \ips$; $\rps -
  \ips, \ips - \zps$; $\ips - \zps, \zps - \yps$).  Left column is
  ubercal data only, right column is all PS1
  data. RA = 0.0 is at the center
  of the plots and increases to the left.} \label{fig:stlocus.midpt}
\end{center}
\end{figure*}

\begin{figure*}[htbp]
\begin{center}
\includegraphics[type=\plotext,ext=.\plotext,read=.\plotext,width=6.0in]{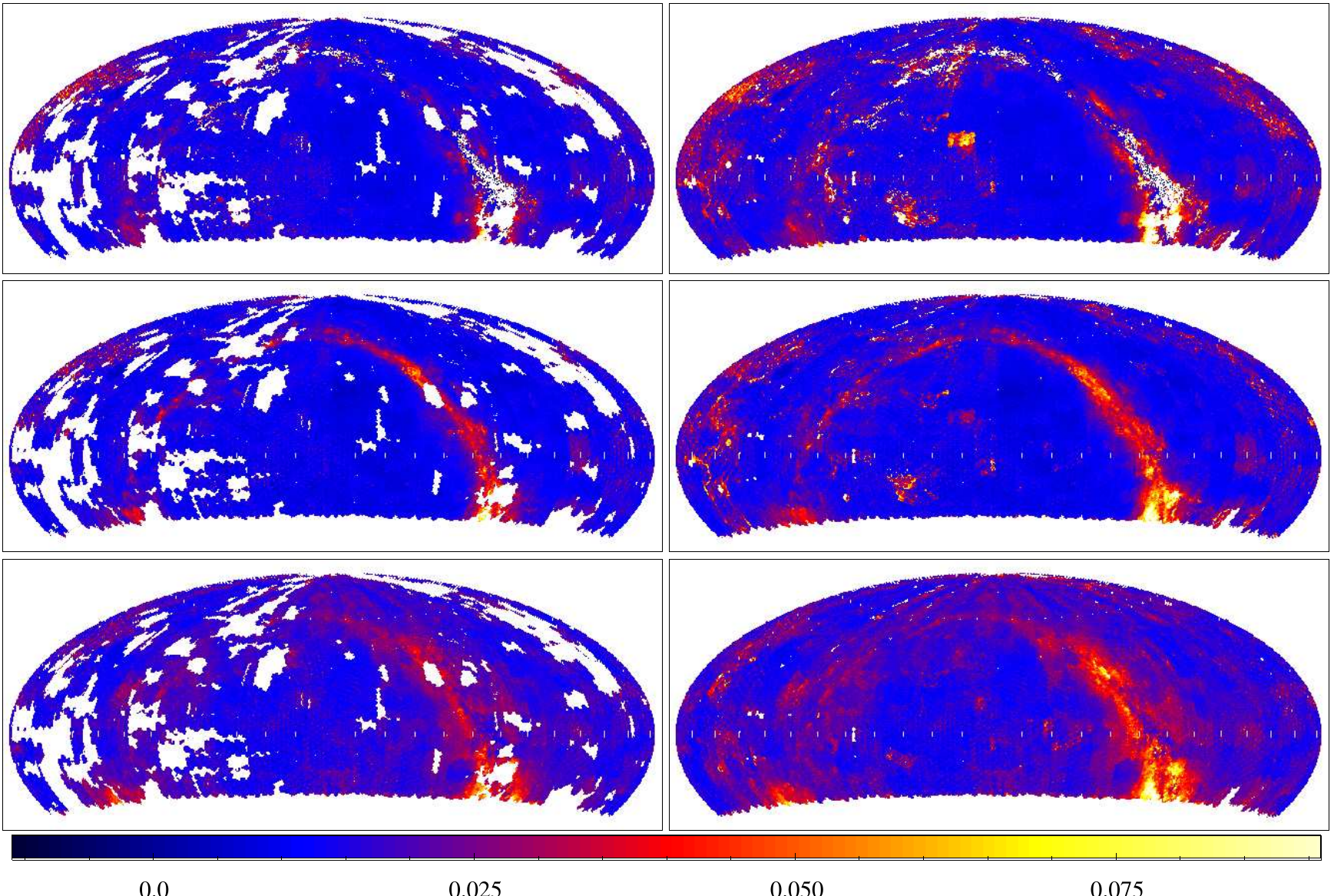}
\caption{Plots of the r.m.s. scatter of the stellar locus as a function
  of position on the sky ($\rps < 20.0$) for the 3 2-color planes
  (top-to-bottom: $\gps - \rps, \rps - \ips$; $\rps - \ips, \ips -
  \zps$; $\ips - \zps, \zps - \yps$).  Left column is ubercal data
  only, right column is all PS1 data. RA = 0.0 is at the center
  of the plots and increases to the left.} \label{fig:stlocus.sigma}
\end{center}
\end{figure*}

Figure~\ref{fig:sigmas} shows the mean scatter for the measurements of
bright stars (instrumental magnitude $< -10$) as a function of
position on the sky (one plot for each filter).  The left panels show
the scatter for only the Ubercal measurements -- these plots
correspond to the figures in \cite{PS1.ubercal}, and illustrate the
level to which we maintain the photometric quality of the Ubercal
analysis.  The right-hand panels show the scatter for all PS1
measurements included in the analysis, through (2011-01-21).  The
scatter is measured by finding the $\pm 1 \sigma$ points on the
cumulative histogram in the assumption of Gaussian statistics -- this
makes the measurement insensitive to extreme outliers, but only if the
outliers are a small fraction of the total measurements.  This figure
shows that the non-photometric data have lower quality than the
photometric data, and as a result the scatter increases for some of
these exposures.

\begin{figure*}[htbp]
\begin{center}
\includegraphics[type=\plotext,ext=.\plotext,read=.\plotext,width=6.5in]{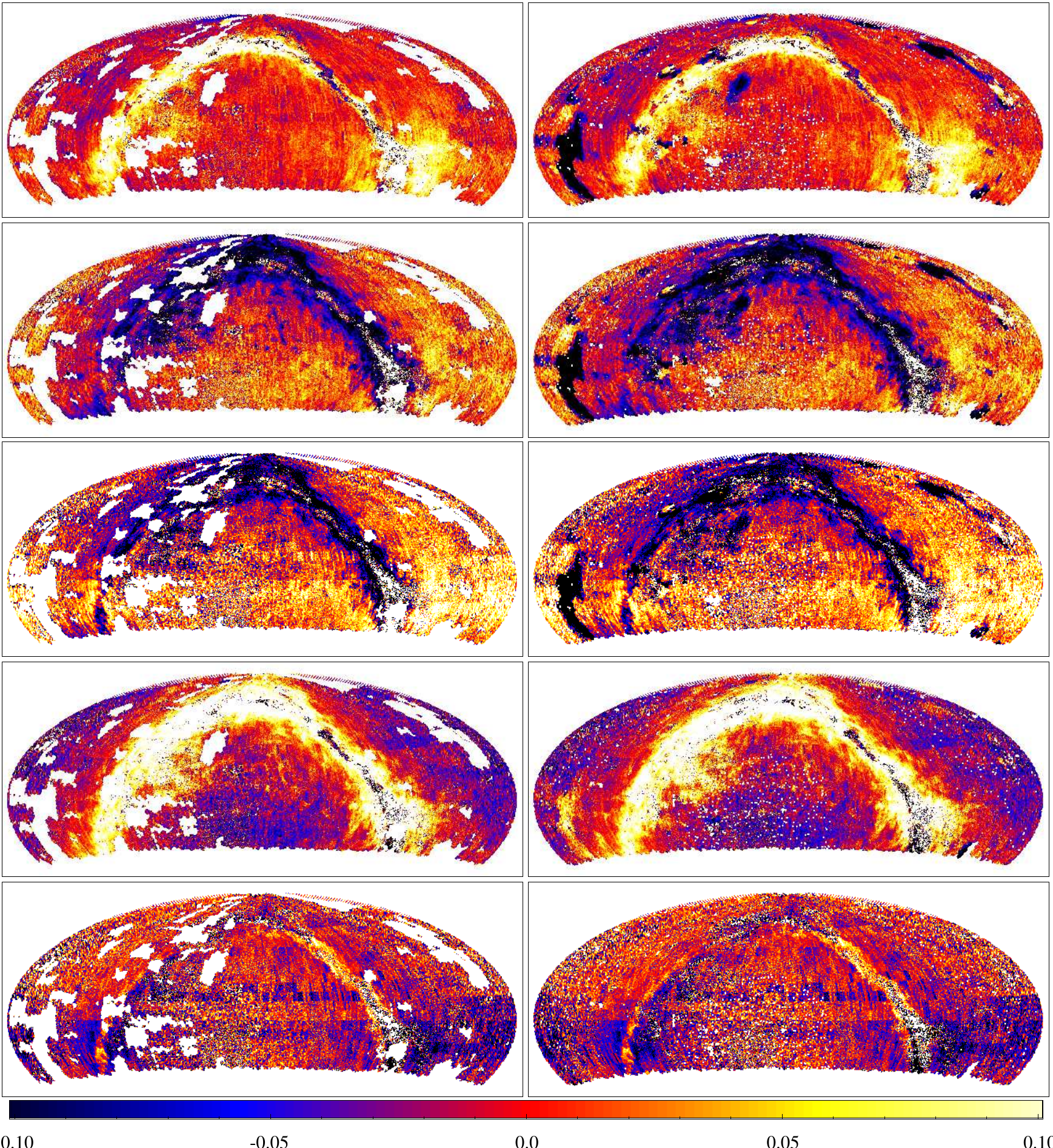}
\caption{Plots of the mean $J,H,K$ offsets and $J - H, H - K$ color
  offsets at fixed $\gps - \ips = 0.5$. Left: Ubercal; Right: all PS1.
  The color scale shows the amplitude of the offset relative to the
  mean of the images. RA = 0.0 is at the center
  of the plots and increases to the left.} \label{fig:2mass}
\end{center}
\end{figure*}

\begin{figure}[htbp]
\begin{center}
\includegraphics[type=\plotext,ext=.\plotext,read=.\plotext,width=3.25in]{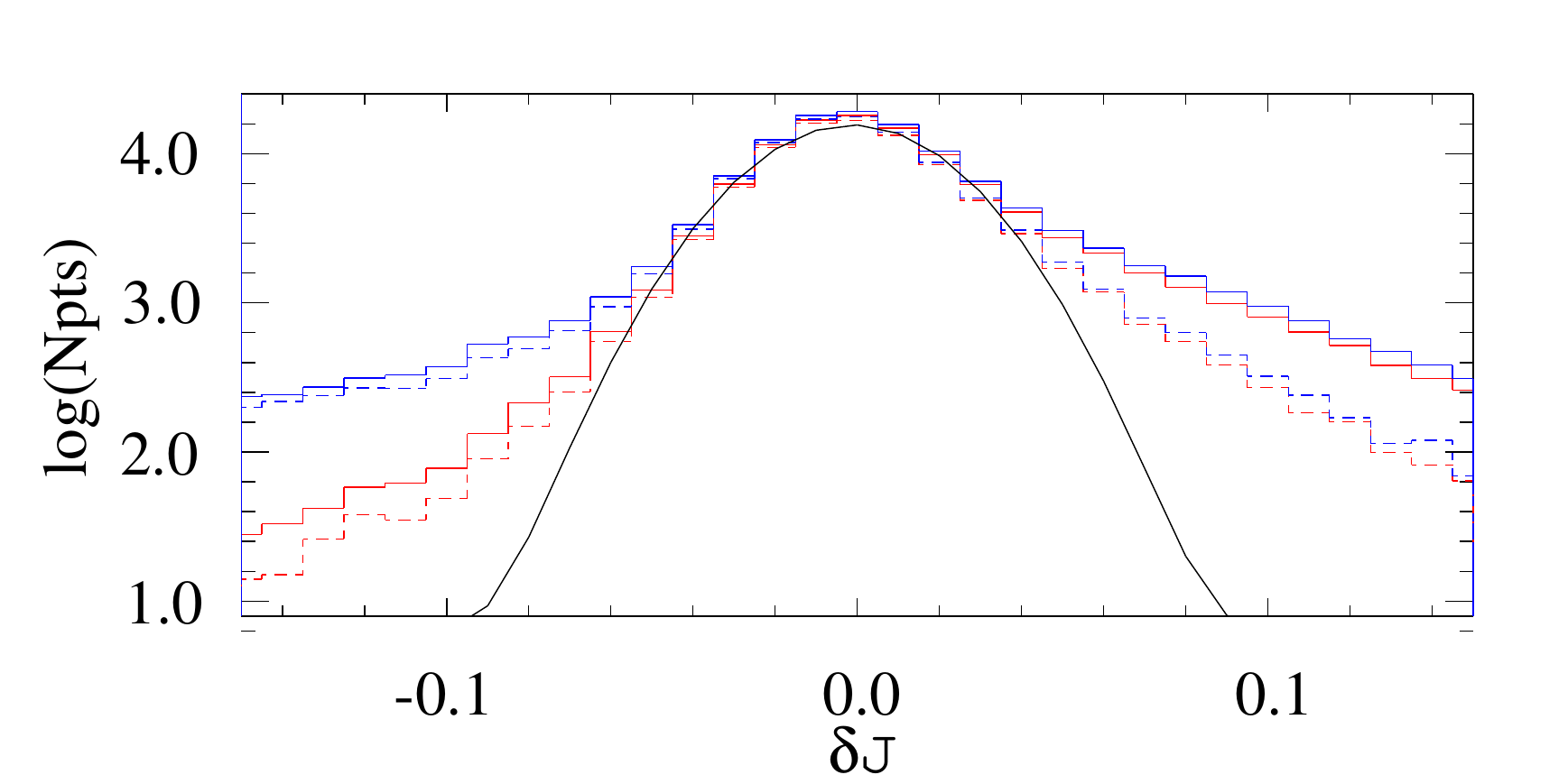}
\caption{Log histogram of $J$ offsets for all (blue lines; thick black
  line in print version) and Ubercal-only (red lines; thin black lines
  in the print version).  Dotted lines show only the regions with $|b|
  > 10.0$\degree.  The black curve is a Gaussian fit to the peak of the
  high Galactic latitude Ubercal-only data ($\sigma =
  0.021$).}\label{fig:2mass-hist}
\end{center}
\end{figure}

From inspection of per-exposure residuals, the cause of the higher
scatter for non-photometric data is clear: with only a single zero
point correction per exposure, we can only correct exposures to a
limited extent.  Those exposures with 2D variations in their effective
extinction show up as trends in the residuals as a function of
position, with corresponding higher scatter.  In detailed examination,
it is clear that a large fraction of these exposures with 2D
attenuation patterns would be well-represented by a linear trend of
the extinction with position.  A future update to the photometric
analysis will attempt to correct such image with a (minimal) 2D trend,
after the full sky photometric coverage is improved.

The other important aspect of Figure~\ref{fig:sigmas} is the improved
coverage resulting from the inclusion of exposures from
non-photometric nights. The gaps in the Ubercal-only plots are filled
in with the non-photometric nights.  Note that in the Ubercal
analysis, images from clearly non-photometric periods, as well as
those from suspect periods, are rejected (either manually or
automatically).  This conservative cut allows for a very clean
solution, but leaves us room to find an overlap solution that tiles
across the gaps with confidence.

\begin{table*}[htdpb]
\caption{\PSONE\ Photometric Ladder Data Fields\label{table:example}}
\begin{center} \tiny
\begin{tabular}{l|r|l}
\hline
\hline
{\bf FITS Name} & {\bf CSV Seq} & {\bf Description} \\
\hline
RA       & 1  & Right Ascension (degrees, J2000) \\
DEC      & 2  & Declination (degrees, J2000) \\
X        & $6 \times N_{band} + 3$ & PSF-fit magnitude for PS1 band (X = \grizy) \\
X:err    & $6 \times N_{band} + 4$ & formal error on magnitude \\
X:nphot  & $6 \times N_{band} + 5$ & number of measurements used for mean magnitude \\
X:stdev  & $6 \times N_{band} + 6$ & standard deviation of mean magnitude \\
X:flags  & $6 \times N_{band} + 7$ & flags for mean magnitude analysis \\
X:ucdist & $6 \times N_{band} + 8$ & distance to Ubercal data in exposure footprints \\
J       & 8  & 2MASS J-band \\
H       & 9  & 2MASS H-band \\
K       & 10 & 2MASS K-band \\
\hline
\multicolumn{3}{l}{X represents one of each filter \grizy}\\
\multicolumn{3}{l}{$N_{band}$ is the filter sequence number \grizy = (0,1,2,3,4)}\\
\hline
\end{tabular}
\end{center}
\end{table*}

For an additional check of the data quality, we employ a technique
similar to the stellar locus analysis described by \cite{Ivezic.2004}
\changed{and \cite{Ivezic.2007}}.  Similar to that analysis, we have selected
segments of the stellar locus in 3 2-color diagrams relevant to the
PS1 data set: $\gps - \rps, \rps - \ips$; $\rps - \ips, \ips - \zps$;
$\ips - \zps, \zps - \yps$ (see Figure~\ref{fig:stlocus.demo}).

We define a reference stellar locus based on the stars in a 400
square-degree region of high Galactic latitude.  We have selected only
bright ($\rps < 20.0$) objects with good measurements in all 5 filters
and no evidence of extendedness.  We have fitted a straight line
segment to the portion of the locus for each 2-color diagram and
determined principal colors ($P1,P2$) along and perpendicular to the
locus in this region.  The linear fits do not describe the stellar
locus extremely well, so we have fitted a spline to the median in 0.05
magnitude wide bins along the $P1$ direction.  We define $Q2$ as the
corrected version of $P2$ with the spline fit subtracted.  Similarly,
there is a small trend of the $P2$ or $Q2$ color as a function of
\rps, so we have again fitted a spline to the median in 0.25 magnitude
wide bins in \rps.  We define $S2$ as the corrected version of $Q2$
with this second spline fit subtracted.  Note that the
\cite{Ivezic.2004} analysis used a linear fit for both color and
magnitude corrections.  The difference between the linear and spline
fits is generally small ($< 0.01$ mag), except for $\ips - \zps, \zps
- \yps$.

The intrinsic width of the stellar locus is quite small, and it is
unclear if we have yet resolved it.  In their analysis of the
$g-r,r-i$ plane, \cite{Ivezic.2004} determine an r.m.s. scatter in the
$P2$ direction of 0.025 mags for single-epoch data, \changed{decreasing to
0.022 mags for 5-epoch data.  \cite{Ivezic.2007} show the $P2$ width
decreasing to 0.010 mags in the many-epoch Stripe 82 data.  Our plot
of $P2$ vs \rps\ (as well as $Q2$ and $S2$) shows a continued
improvement to brighter objects, with an observed width of 0.012
magnitudes for $\rps < 16.0$ (see Figure~\ref{fig:stlocus.demo}),
quite comparable to the Stripe 82 observations from SDSS.}

We have fitted the stellar locus defined above (for each 2-color
plane) to the positions of stars for 0.5\degree boxes across the sky,
using the same restrictions on magnitude range and quality when
selecting the stars. In Figure~\ref{fig:stlocus.midpt}, we present the
mean offset of the observed stellar locus relative to the template for
stars of $\rps < 20.0$.  In Figure~\ref{fig:stlocus.sigma}, we present
the scatter of the stellar locus for \rps\ $< 20.0$.  Again this shows
the high quality of the data using the ubercal analysis, but the
limited coverage compared to the full \TPS\ area.  These figures also
illustrate that our relative photometric analysis can extend the
regions of good coverage, but that the quality is sometimes degraded
excessively.  

\begin{figure*}[htbp]
\begin{center}
\includegraphics[type=\plotext,ext=.\plotext,read=.\plotext,width=6.5in]{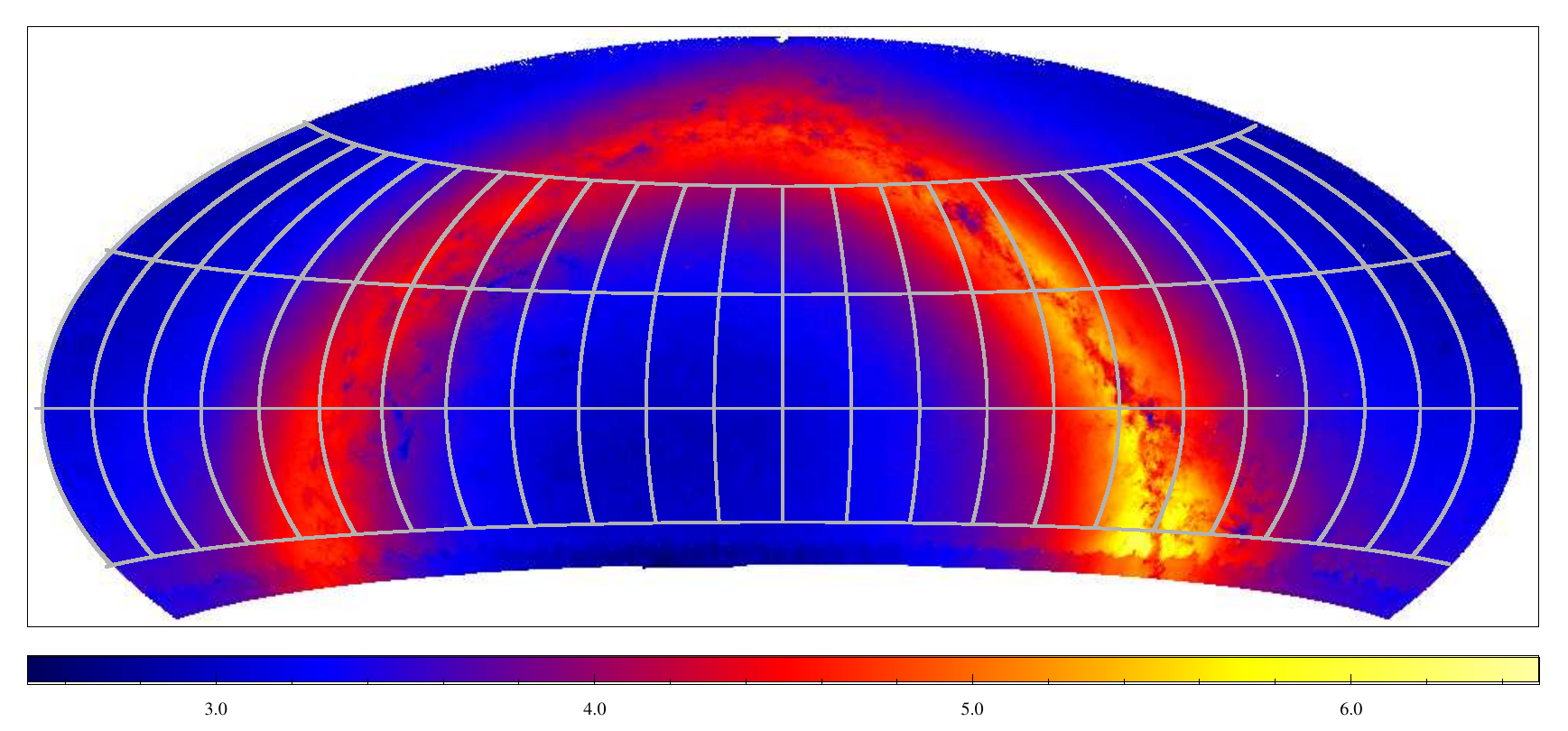}
\caption{Location of the photometric ladder overlayed on a plot of the
  spatial density of objects.  The color scale gives the logarithm of
  the number of objects per square degree with at least 3 measurements
  and $\rps < 19.0$. RA = 0.0 is at the center
  of the plots and increases to the left.} \label{fig:ladder}
\end{center}
\end{figure*}

\subsubsection{Comparisons with 2MASS}

To explore how well relative photometry allows us to patch across the
empty areas, we have compared our data to 2MASS observations, the only
photometric dataset approaching our data quality and coverage (SDSS
has large gaps in the regions of interest).  For each
0.5\degree\ pixel, we select the objects with high quality photometry
(\gps, \rps, \ips\ errors $< 0.02$, standard deviations $< 0.05$
magnitudes), excluding objects thought to be extended in these bands
(based on PSF vs aperture photometry).  We also exclude saturated
objects. We then generate a color-color diagram in $\gps - \rps, \rps -
\ips$ and select the objects in the color range $0.2 < \gps - \rps < 1.0$, and
within 0.05 magnitudes of the stellar locus in $\gps - \ips$.  The resulting
sample should be dominated by early type stars for which the optical /
IR color-color loci are well defined.

We fit the color-color locus for the selected stars in each of the
optical / IR color-color diagrams: $\gps - \ips, \ips - J$; $\gps -
\ips, \ips - H$; $\gps - \ips, \ips - K$. Finally, we determine
`color offsets', the fitted values of $\ips - J,H,K$ at a fixed
$\gps - \ips$ value of 0.5.  These color offsets are determined for each
of the 0.5\degree\ patches across the sky.  If the data consisted of
only stars with well-measured magnitudes and low extinction, the color
offsets should be quite consistent across the sky.  Extinction will
shift the color-color loci, as well any photometric errors in either
PS1 or 2MASS.  At a lower level, changes in the mean metallicity or
surface gravity of the relevant population can affect the color
offsets as well.  In Figure~\ref{fig:2mass}, we present maps of the
color-offsets across the full $3\pi$ region.  The left panels show
only the Ubercal-tied data (\gps, \rps, \ips\ all required), while the
right panels show all data.  The first three panels show the color
offsets for $J,H,K$ respectively, while the last two panels show
$J - H$ and $H - K$.

Several features can be seen in these plots.  First, the Galactic
Plane is clearly visible in all filters and combinations.  This
reflects the extinction sensitivity of these measurements.  Further
investigation of the 2D (and 3D) extinction patterns is ongoing and
will be a major data product of the \PSONE\ \TPS\ \citep{PS1.3D.dust}.

Next, large scale structures can be seen which generally correspond to
areas without {\em Ubercal} calibrations.  This illustrates the
difficulty of using relative photometry alone to tile across very
large patches.  While there are regions with acceptable consistency
which lack {\em Ubercal} data, it is clear that the best quality
photometry will only be possible when photometric data are available for
essentially the whole survey region.  Figure~\ref{fig:2mass-hist}
shows the impact of {\em Ubercal} on the photometric residuals.  In
this figure, the blue line is a (log) histogram of the $\ips-J$
offsets using all data, while the red line shows the same using only
the {\em Ubercal} measurements.  The black line is a Gaussian fit to
the core of the distribution, with $\sigma = 0.023$ magnitudes.  The
positive outliers in this histogram come principally from the Galactic
Plane, while the low end outliers show the impact of calibration
errors on the non-{\em Ubercal} analysis.

Finally, there are visible patterns unrelated to any of the Pan-STARRS
1 observing pattern or sky tessellation.  The strongest of these is
the banding pattern seen in $K$ and $H-K$ running from RA $\sim 290$
to $\sim 90$, with boundaries at declinations of roughly 0.0, 6.0,
12.0, and 18.0\degree.  At a lower level, a stripping pattern of
narrow North-South bands a few RA degrees wide can also be seen.  We
believe these can be attributed to calibration offsets in the 2MASS
data, especially in $K$ band.  The amplitude of these patterns is
consistent with a standard deviation of 2 - 3\% for these regions.

This analysis, and comparisons between PS1 and SDSS
\citep{PS1.MD.stds} illustrate the difficulty of producing a single
consistent photometric data product on a large scale.  Especially for
2MASS and SDSS, the calibration is made additionally challenging by
virtue of having only a single epoch for most of the survey area.  The
lack of internal repeat checks make it very difficult to prove that
the calibrations are accurate across the whole survey region.

Even with multiple epochs, PS1 photometry calibration still has room
for improvement, as can be seen in the figures above.  However, as the
PS1 survey is on-going, our expectation is that we will be able to
fill in the non-photometric gaps in the upcoming years.  The full
power of the large-area and all sky surveys can be realized by
combining the dataset to constrain and correct the systematics from
all of these surveys.  

\section{DATA PRODUCTS}

\begin{table*}[ht]
\caption{Bit-flags used to characterize average photometry values} 
\begin{center} \tiny
\begin{tabular}{l|l|l}
\hline
\hline
{\bf FLAG NAME} & {\bf Hex Value} & {\bf Description} \\
 ID\_SECF\_USE\_SYNTH   & 0x00000004 & synthetic photometry used in average measurement \\
 ID\_SECF\_USE\_UBERCAL & 0x00000008 & Ubercal photometry used for average measurement \\
 ID\_PHOTOM\_PASS\_0    & 0x00000100 & pass 0 : only use measurements thought to be GOOD (based on photometry analysis) \\
 ID\_PHOTOM\_PASS\_1    & 0x00000200 & pass 1 : accept measurements thought to be POOR (based on photometry analysis) \\
 ID\_PHOTOM\_PASS\_2    & 0x00000400 & pass 2 : accept the measurements marked as outliers based on relative photometry \\
 ID\_PHOTOM\_PASS\_3    & 0x00000800 & pass 3 : accept measurements thought to be BAD (based on photometry analysis) \\
 ID\_PHOTOM\_PASS\_4    & 0x00001000 & pass 4 : accept the measurements outside of the instrumental magnitude limits (eg, SAT) \\
 ID\_SECF\_OBJ\_EXT     & 0x01000000 & possibly extended in this band \\
\hline
\end{tabular}
\end{center}
\label{table:flags}
\end{table*}

Over the next 1.5 - 2 years, Pan-STARRS 1 will finish the initial
basic survey, completing coverage of the full $3\pi$ region.  One of
our major eventual goals is the production of a high-quality
astrometric and photometric reference catalogs which can be used by
observers to calibrate data with in-field references.  The full survey
dataset is necessary to pin down all regions at the highest possible
precision.  The vagaries of weather means that some substantial
patches have not yet been observed in photometric conditions, and are
only currently tied via relative photometry to the rest of the survey.

In this initial public release, we are providing the PS1 Reference
Photometry Ladder as a sample of future \PSONE\ photometric
calibration data.  The Ladder consists of 4 strips in right ascension,
1 degree high, at declinations of (-25, 0, +25, +50), and 24 strips in
declination, 1 RA degree wide centered on RA hours (see Figure~\ref{fig:ladder}).


We have selected a subset of bright ($\gps, \rps, \ips < 19.14, 19.26,
19.24$), well-measured ($> 4$ observations total) objects in these
regions.  The brightness limit was chosen to be equivalent to an
instrumental magnitude of -9.5, or roughly 2\% statistical errors.
This brightness cut not only restricts the sample to objects with high
signal-to-noise, it also avoids the stars affected by inconsistencies
for faint stars.  The inconsistencies we have observed take the form
of variations in the measured flux for faint stars, relative to bright
stars in the same images, as a function of seeing.  We have also
limited the density to $\sim 1000$ objects for each 1 square degree
patch.  To choose a high-quality sample at this limited density, we
used the mean $\chi^2$ value for the \gps, \rps, and \ips\ mean
magnitudes to select the high-quality measurements.  We then selected
the 1000 objects for a given patch in ascending order of $\chi^2$ so
that objects with consistent photometry are preferred.

The data are available from the PS1 Public Data web
site\footnote{http://ipp.ifa.hawaii.edu}.  

\changed{We provide two sets of photometry tables: ``Ubercal'' and
  ``Relphot''.  The ubercal tables only include objects for which all
  5 filters have measurements calibrated directly in the ubercal
  analysis.  The relphot tables accept objects for which some filters
  have been tied to the ubercal data via relative photometry.  As
  discussed above, the relphot values have lower confidence and should
  not be used for precision photometry.  However, the relphot tables
  provide additional sky coverage.  In most of these areas, a subset
  of the filters have ubercal photometry.  Careful use of the
  per-filter flags is advised!}

The tables are provided as {\tt FITS} and comma-separated value
tables.  Table~\ref{table:example} defines the columns available in
the tables.  There are 27 separate tables, 1 for each of the ladder
rungs in Dec and 1 for each of the rungs in RA.  Note that there are
duplicate entries between files where the rungs overlap.

Table~\ref{table:example} describes the fields provided for each
object.  Data for each of the five filters (\grizy) are given in
blocks, with 6 parameters provided for each filter.  The magnitude
provided is the mean PSF magnitude calibrated as described above.  The
error is the formal error determined by accepting the reported error
from the photometric analysis for each separate measurement.
$X:nphot$ represents the number of measurements actually used to
determine the mean magnitude.  Because of outlier clipping, this
number may be smaller than the number of separate epochs available for
a given object.  The standard deviation of the mean magnitude is
reported {\em after} clipping has been applied.  The mean magnitude
calculated for each filter is coupled to a set of bit-flags which
describe the averaging process (Table~\ref{table:flags}).  This value is reported as a 32-bit
integer, but only the following 8 bits are used for each filter (see
the discussion in Section~\ref{sec.relphot}).  

To assess the extendedness of an object, we use the difference between
the PSF and aperture magnitudes.  If this difference is larger than
0.1 added in quadrature to 2.5 times the statistical error, then the
measurement is considered extended.  This measurement shows the
presence of additional light in the aperture beyond the core
consistent with the PSF model.  The flag noting a possible extended
object is raised if more than 50\% of the measurements in a given band
are observed to be extended based on this test.


{\it Facilities:} \facility{PS1 (GPC1)}

\acknowledgments

The \PSONE\ Survey has been made possible through contributions of the
Institute for Astronomy, the University of Hawaii, the Pan-STARRS
Project Office, the Max-Planck Society and its participating
institutes, the Max Planck Institute for Astronomy, Heidelberg and the
Max Planck Institute for Extraterrestrial Physics, Garching, The Johns
Hopkins University, Durham University, the University of Edinburgh,
Queen's University Belfast, the Harvard-Smithsonian Center for
Astrophysics, and the Las Cumbres Observatory Global Telescope
Network, Incorporated, the National Central University of Taiwan, and
the National Aeronautics and Space Administration under Grant
No. NNX08AR22G issued through the Planetary Science Division of the
NASA Science Mission Directorate. Partial support for this work was
provided by National Science Foundation grants AST-1009749 \& AST-1238877.



\begin{thebibliography}


\bibitem[Anderson, G.P.,~{\it et al.}(2001)]{MODTRAN} Anderson, G.~P., {\it et al.} 2001, \procspie\ 4381, 455


\bibitem[Bessel(1990)]{Bessel.1990} Bessel, M.~S.\ 1990, \pasp{102}{1181}

\bibitem[Chambers {\it et al.}(in prep)]{PS.MDRM} Chambers, K.~C.,  {\it et al.}, in preparation.

\bibitem[Finkbeiner {\it et al.}(in prep)]{PS1.MD.stds} Finkbeiner, D.~P., {\it et al.}, in preparation.

\bibitem[Hodapp {\it et al.}(2004)]{PS1.optics} Hodapp, K.~W., Siegmund, 
W.~A., Kaiser, N., Chambers, K.~C., Laux, U., Morgan, J., 
\& Mannery, E.\ 2004, \procspie\ 5489, 667 

\bibitem[Hsieh {\it et al.}(2012)]{Hsieh-2012} Hsieh, H.~H., Yang, B., Haghighipour, N., Kaluna, H.~M., Fitzsimmons, A., {\it et al.}, 2012, Asteroids, Comets, Meteors 2012, Proceedings of the conference held May 16-20, 2012 in Niigata, Japan. LPI Contribution No. 1667, id.6313.

\bibitem[Ivezi\'c {\it et al.}(2004)]{Ivezic.2004} Ivezi\'c, Z.,
  Lupton, R.~H., Schlegel D., {\it et al.,} 2004, Astronomische
  Nachrichten 325, 583.

\bibitem[Ivezi\'c {\it et al.}(2007)]{Ivezic.2007} Ivezi\'c, Z.,
  Smith, A., Miknaitis G., {\it et al.,} 2007, \aj{134}{973}

\bibitem[Johnson {\it et al.}(2009)]{Johnson.2009} Johnson, J.~A.,
  Winn, J.~N., Cabrera, N.~E., Carter, J.~A.\ (2009) \apjl{692}{100}

\bibitem[Kaiser {\it et al.}(2010)]{PS1.system} Kaiser, N., {\it et al.} 2010, \procspie\ 7733, 12K.

\bibitem[Magnier(2006)]{PS1.IPP} Magnier, E.\ 2006, Proceedings of The Advanced 
Maui Optical and Space Surveillance Technologies Conference, Ed.: S. Ryan, The Maui Economic Development Board, p.E5



\bibitem[Narayan et al.(2011)]{2011ApJ...731L..11N} Narayan, G., Foley, R.~J., Berger, E., et al.\ 2011, \apjl{731}{11}

\bibitem[Onaka {\it et al.}(2008)]{PS1.GPCB} Onaka, P., Tonry, J.~L., 
Isani, S., Lee, A., Uyeshiro, R., Rae, C., Robertson, L., 
\& Ching, G.\ 2008, \procspie\ 7014, 12O

\bibitem[Padmanabhan et al.(2008)]{SDSS.Padmanabhan} Padmanabhan, N., Schlegel, D.~J., Finkbeiner, D.~P., et al.\ 2008, \apj{674}{1217}


\bibitem[Tonry \& Onaka(2009)]{PS1.GPCA} Tonry, J., \& Onaka, P.\ 2009, Advanced Maui Optical and Space Surveillance Technologies Conference, 
Proceedings of the Advanced Maui Optical and Space Surveillance Technologies Conference, Ed.: S. Ryan, p.E40.   

\bibitem[Tonry {\it et al.}(2005)]{Tonry.2005} Tonry, J.~L., Howell,
  S.~B., Everett, M.~E., Rodney, S.A., Willman, M., VanOutryve,
  C.\ 2005, \pasp{117}{218}

\bibitem[Tonry {\it et al.}(2012)]{JTphoto} 
Tonry, J.~L., Stubbs, C.~W., Lykke, K.~R., Doherty, P., Shivvers, I.~S., Burgett, W.~S., Chambers, K.~C., Hodapp, K.~W., Kaiser, N., Kudritzki, R.-P., Magnier, E.~A., Morgan, J.~S., Price, P.~A., Wainscoat, R.~J.
2012, \apj{750}{99}

\bibitem[Schlafly {\it et al.}(2012)]{PS1.ubercal} Schlafly,
  E.~F., Finkbeiner, D.~P., Juri\'c, M., Magnier, E.~A., Burgett,
  W.~S., Chambers, K.~C., Grav, T., Hodapp, K.~W., Kaiser, N.,
  Kudritzki, R.-P., Martin, N.~F., Morgan J.~S., Price, P.~A., Stubbs,
  C.~W., Tonry, J.~L., Wainscoat, R.~J. 2012, \apj{756}{158}.

\bibitem[Schlafly {\it et al.}(in prep)]{PS1.3D.dust} Schlafly, E.~F., et al. 2012, in preparation

\bibitem[York {\it et al.}(2000)]{SDSS.York} York, D.~G., Adelman, J.,
  Anderson Jr., J.~E., Anderson, S.~F., Annis, J., Bahcall, N.~A.,
  Bakken, J.~A., Barkhouser, R., Bastian, S., Berman, E., Boroski,
  W.~N., Bracker, S., {\it et al.,} 2000, \aj{120}{1579}.

\end{thebibliography}
\end{document}